# Quantifying stored energy release in irradiated YBa$_2$Cu$_3$O$_7$ through molecular dynamics annealing simulations

L. D. Kortman[1,3], A. R. Devitre[2], C. A. Hirst[3,4]

[1] Department of Materials Science and Engineering, Massachusetts Institute of Technology, Cambridge, USA
[2] Department of Nuclear Science and Engineering, Massachusetts Institute of Technology, Cambridge, USA
[3] Department of Nuclear Engineering and Radiological Sciences, University of Michigan, Ann Arbor, USA
[4] Department of Nuclear Engineering and Engineering Physics, University of Wisconsin, Madison, USA

E-mail: `kortman@umich.edu, cahirst@wisc.edu`

22 May 2025

**Abstract.**
   Over the lifetime of a fusion power plant, irradiation-induced defects will accumulate in the superconducting magnets compromising their ability to carry current without losses, generate high magnetic fields, and thus maintain plasma confinement. These defects also store potential energy within the crystalline lattice of materials, which can be released upon annealing. This phenomenon raises the question of whether the energy stored in defects may be sufficient to accelerate, or even trigger, a magnet quench? To provide an order of magnitude estimate, we used molecular dynamics simulations to generate defected YBCO supercells and conduct isothermal annealing simulations. Our results reveal that the maximum volumetric stored energy in a 4 mDPA defected single crystal of YBCO (240 $J/cm^3$) is 30 times greater than the experimental minimum quench energy values for YBCO tapes (8.1 $J/cm^3$) [1]. Our simulations also show that the amount of energy released increases as a function of annealing temperature or irradiation dose. This trend demonstrates that localized heating events in an irradiated fusion magnet have the potential to release significant amounts of defect energy. These findings underscore the critical need for experimental validation of the accumulation and release of defect stored energy, and highlight the importance of incorporating this contribution into quench detection systems, to enhance the operational safety of large-scale YBCO fusion magnets.

## 1. Introduction

Fusion reactors can be built smaller, faster, and cheaper by using superconducting magnets made of rare-earth barium copper oxide (REBCO) materials [2]. REBCO's higher critical current, compared to conventional superconductors e.g. Nb$_3$Sn or NbTi, enables generation of stronger magnetic fields, improving plasma confinement and allowing for more compact reactor designs [3, 4].

The operation of a D-T fusion reactor produces high-energy neutrons that collide with the atoms inside reactor components, leaving behind a variety of irradiation-induced defects such as vacancies, interstitials, and their clusters. Over time, the accumulation of defects has a net detrimental effect on component performance; the superconducting magnets, for example, gradually lose their ability to carry current without losses – also known as the critical current ($I_c$) – which compromises their ability to generate the magnetic field that sustains fusion reactions in the core [5].

Controlled annealing of the magnets, for instance during maintenance periods, can prolong the operational lifetime of fusion power plants (FPP). By intentionally increasing the magnet temperature, the mobility of defects increases, promoting their rearrangement within the crystal lattice. For example, interstitials can find each other and cluster, or find a vacancy and annihilate [6]. Through this process, a fraction of $I_c$ is recovered by eliminating scattering centers [7], and the superconducting magnets' lifetime is extended.

Uncontrolled annealing, on the other hand, can lead to a thermal runaway event known as a *magnet quench*. Multiple scenarios can initiate a quench, e.g. a sudden loss of coolant, localized resistance from manufacturing defects in the YBCO layer, or mechanical strain imposed on the superconductor. These conditions can lead to a local transition out of the loss free state [8]. A high current running through a (now) resistive conductor will increase the temperature, which increases the resistance and creates a positive feedback thermal runaway event.

By rearranging themselves into lower-energy configurations, irradiation-induced defects can release the potential energy that was invested in their formation in the form of heat [9]. As such, the release of energy stored in defects could cause local heating during operation and could potentially accelerate, or initiate, a magnet quench. The relationship between irradiation dose, defect density, stored energy, irradiation temperature, and defect energy release must therefore be quantified to evaluate the likelihood of a quench following the accumulation of irradiation-induced defects in YBCO magnets.

Superconducting REBCO magnets are made of coated conductors, or tapes, in which most of the mass is comprised of copper and Hastelloy [10]. While REBCO typically amounts to less than 5% of the tape mass, this study focuses on the energy release from defects in this layer, since irradiated ceramics can store 100–1,000 times more energy than metals [11] and YBCO carries most of the current when a quench starts. Here we focus on yttrium barium copper oxide (YBa$_2$Cu$_3$O$_7$) as one of the front-running REBCO variants considered for fusion magnets [12, 13]. An additional clarification is made to distinguish between the energy stored in crystalline defects *before* a quench, and the *magnetic* stored energy that has the potential to be released *following* a magnet quench.

The energy needed to initiate a magnet quench is known as the minimum quench energy (MQE). The MQE has been measured for single tapes [14] and cables [8], but its exact value depends on the precise measurement conditions including test geometry, specific materials, and cooling capacity. While quenches can propagate in multiple directions, experiments typically focus on quench propagation along the length of the superconductor. In these setups, a heater is typically placed at the center of a YBCO tape. Thermocouples and voltage taps are spaced down the length of the tape to detect the magnitude and velocity of a quench [8, 14, 15]. We have adopted MQE as a comparison metric for this study, to provide context for the defect energies calculated by our simulations; it should be noted that these parameters are not directly comparable due to the many differences in scale, structure, and context, which is explored further in the discussion.

The loss of the superconducting state in YBCO can occur from a combination of operational and irradiation-induced stressors. During operation, exceeding a certain temperature, current, or external magnetic field will transition the superconductor out of the superconducting regime, these parameters are subsequently designated as the critical temperature ($T_c$), critical current ($I_c$), critical current density ($J_c$) and critical magnetic field ($B_c$) [16]. Energy imparted on the lattice from incoming particles has been directly linked to $T_c$ suppression in YBCO [17]. Thermal loads can be driven by neutrons and gammas passing through the material, where energy is deposited into the lattice due to the deceleration of particles [18]. The energy released by defect annihilation is an additional contributor to localized thermal loads, and needs to be explored in the context of magnet operation. Rather than the direct measurement of defect energetics, prior experiments on YBCO have primarily focused on the degradation of superconducting properties ($I_c$, $T_c$) due to defect generation from irradiation [5, 7, 19]. Fast neutron irradiation has been shown to initially enhance flux pinning at low fluences but leads to performance degradation beyond a critical defect density, with the threshold highly dependent on temperature [5]. Transmission electron microscopy (TEM) studies have directly observed defects approximately 2–5 nm in size and indicated the formation of amorphous regions, strain fields, and complex microstructures at higher doses [19]. Furthermore, annealing experiments demonstrated partial recovery of $I_c$ and $T_c$, consistent with defect recombination and reordering processes [7].

Quantifying the stored energy in irradiation-induced defects is difficult to do experimentally due to the cryogenic operating temperatures required for representative irradiation conditions, and the need to maintain this temperature before atomic-scale characterization of the created defects. As an alternative strategy, computer simulations can be used to estimate the volumetric energy released by defects. Molecular dynamics (MD) is an atomistic simulation method that computes the motion of atoms and can

4be used to model the evolution of defects in YBCO.

The first known MD simulations of radiation damage in YBCO were performed by Cui et al. [20] in 1992, using the potential developed by Chaplot [21]. This interatomic potential was later extended by Murphy et al. [22] to allow Cu ions to switch between two unique sites, and was used to demonstrate the creation of amorphous regions resulting from the recoil of 2–5 keV Barium primary knock-on atoms in the lattice. While MD annealing simulations of various ceramics have been documented in the literature [23,24], to the best of our knowledge, YBCO MD annealing simulations have not been conducted previously.

## 2. Methods

This project investigated the energy stored in YBCO defects through MD simulations, using the August 2$^{nd}$, 2023 version of LAMMPS [25] and the interatomic potentials developed by Chaplot [21] and Murphy et al. [22]. The Chaplot potential was developed in 1989 using empirical fitting to reproduce structural and vibrational properties, while the Murphy potential was developed in 2020 via density functional theory (DFT) calculations. In this study, initially, both potentials were used to compute the energy stored by defect concentrations ranging from 0 to 0.8 at.% at a temperature of 0 K. This range of defect concentrations was determined based on experimental studies of YBCO irradiated with 25 MeV oxygen ions between 60 K–80 K, which showed that the critical temperature degrades between doses of 1 and 9 mDPA [26]. Defect concentrations of 0–0.8 at.% correspond to canonical doses of 0–8 mDPA (milli-displacements per atom), calculated as the ratio of the number of inserted Frenkel pairs to the total number of atoms [27]. For comparison, the average lifetime dose of YBCO in designs similar to Commonwealth Fusion System's ARC reactor is estimated to be 4 mDPA, but a dose of 7.2 mDPA may be reached in a full power year for magnet regions near the divertor [28].

A YBCO supercell of 40×40×16 unit cells (332,800 atoms) was used for simulations with the Chaplot potential, and a larger supercell of 60×60×20 unit cells (936,000 atoms) was used for the Murphy potential. Both supercells employed periodic boundary conditions to create infinite single crystals. To determine the energy stored in defects at 0 K, Frenkel pairs were created within each supercell by using the *displace_atoms* LAMMPS command. This command displaces a random atom by a random (X, Y, Z) vector, with randomness controlled by a pseudo-random number generator. Different seed numbers were used to introduce variation between simulations. The displacement process is conducted without modeling atomic trajectories [29] and iterates to create a chosen number of defects. This method of defect insertion – known as Frenkel pair accumulation (FPA) algorithm – has previously been used to study primary radiation damage in tungsten [27, 30], zirconium [31], and iron [27, 32]. Strictly speaking, the FPA algorithm [27] most closely represents electron irradiation at 0 K, and it does not account for dose rate effects that influence defect evolution at non-zero temperatures. It is also noted that the primary radiation damage caused by neutrons will differ from that generated by the FPA algorithm, and further work includes investigating defect accumulation following multiple consecutive primary knock-on atom (PKA) cascades [29, 30]. The potential energy of the pristine cell was subtracted from the potential energy of a defected cell, to determine the energy stored in crystalline defects. After comparing the results using both potentials, the Murphy potential was used exclusively for the remainder of the simulations, as it represents the most up-to-date interatomic potential for YBCO.

The series of simulations used to prepare supercells for evaluation is demonstrated in Figure 1. First, a pristine supercell is held at 1 K for 10 ps, then over 50 ps it is heated to the desired temperature (1 K, 20 K, 40 K, 92 K, and 300 K) using the

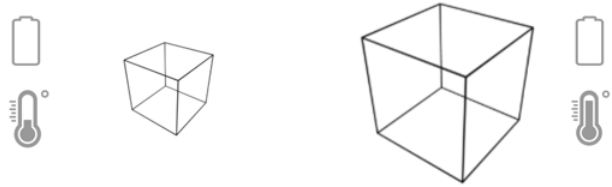

1. Expand cell: NPT simulation 1K to temperature

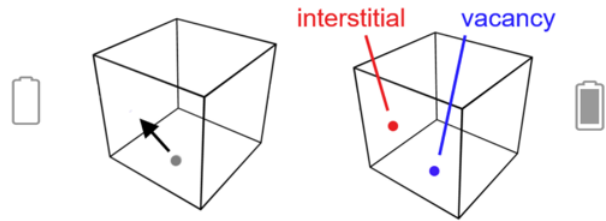

2. Insert defects: Frenkel pair accumulation

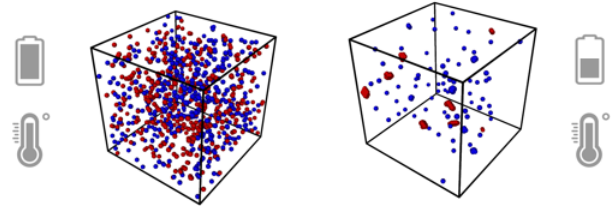

3. Anneal: NVT hold at temperature

**Figure 1. To study defect accumulation and recovery, simulation cells are heated, atoms are displaced, and the cells are relaxed at temperature.** Cells were heated to temperatures between 1 K and 300 K over 60 ps. 468–7488 atoms were displaced using the Frenkel pair accumulation method, resulting in canonical doses of 0.5–8 mDPA. Finally, cells were held isothermally for 35 ps. Note that perfect atoms are not shown in the figure.

constant number of atoms, pressure, and temperature (NPT) ensemble, to allow for thermal expansion. Next, the FPA algorithm is used to create defects in the expanded cell. In order to correct for any artifacts generated by the defect production process, supercells were subsequently held isothermally for 35 ps. Of the chosen temperatures, those of most practical relevance include 20 K, the proposed operating temperature for YBCO in fusion reactors [8]; 92 K, the temperature above which optimally doped YBCO crystals become resistive [33]; and 300 K, room temperature. Local hot spots within a YBCO tape can reach hundreds of Kelvin, and local temperatures up to 300 K do not cause permanent damage to YBCO performance [34]. During isothermal annealing, a timestep of 1 fs and the canonical NVT ensemble [35] were used. The potential energy of each atom was calculated post-isotherm using the LAMMPS *thermo_style pe* command and all atoms were summed together. The potential energy output, in eV, is converted to J/cm$^3$ using Equation 1, where V is the simulation cell volume in angstroms.

$$\mathrm{E}_{stored}\left[\frac{J}{cm^3}\right] = \frac{E_{stored}\,[eV]}{936{,}000\,[atoms]} \times \frac{936{,}000\,[atoms]}{V\,[\mathring{A}^3]} \times \frac{10^{24}\,[\mathring{A}^3]}{1\,[cm^3]} \times \frac{1.602\times10^{-19}\,[J]}{1\,[eV]}$$

Next, the defect stored energy is calculated as a function of time, temperature, and DPA according to the following definition:

$$E^T_{stored}(t) = E^T_{mDPA}(t) - E^T_{0mDPA}(t) \tag{1}$$

where $E^T_{stored}(t)$ is the energy contribution from defects at a given temperature (T), time (t), or dose (mDPA). $E^T_{mDPA}(t)$ is the potential energy of the supercell at a given dose, temperature, and time. $E^T_{0mDPA}(t)$ is the potential energy of the pristine supercell at the same temperature and time. The energy difference between the defected cell and the pristine cell at the same temperature and simulation time is defined as the energy *stored* by defects.

The energy *released* by defects is similarly calculated by subtracting the final stored energy from the initial stored energy at a given temperature or time:

$$E^T_{released}(t) = E^T_{0mDPA}(t) - E^T_{mDPA}(t) - E^{T_0}_{mDPA}(t_0) \tag{2}$$

such that $E^T_{released}(t)$ is the cumulative energy released between the initial time $(t_0)$ and the current timestep $(t)$, or the cumulative energy released between the initial temperature $(T_0)$ and the current temperature $(T)$.

## 3. Results

At 0 K, the energy stored in defects is at a maximum due to the inability for defects to migrate and recombine. By evaluating the energy stored in defects at 0 K, we have determined an upper bound on the potential energy that defects can store in the material. The energy stored in defects at 0 K was evaluated across different damage levels using FPA on pristine cells, inserting defects up to 8 mDPA. Figure 2 compares the stored energy as a function of irradiation dose between the Chaplot and Murphy interatomic potentials.

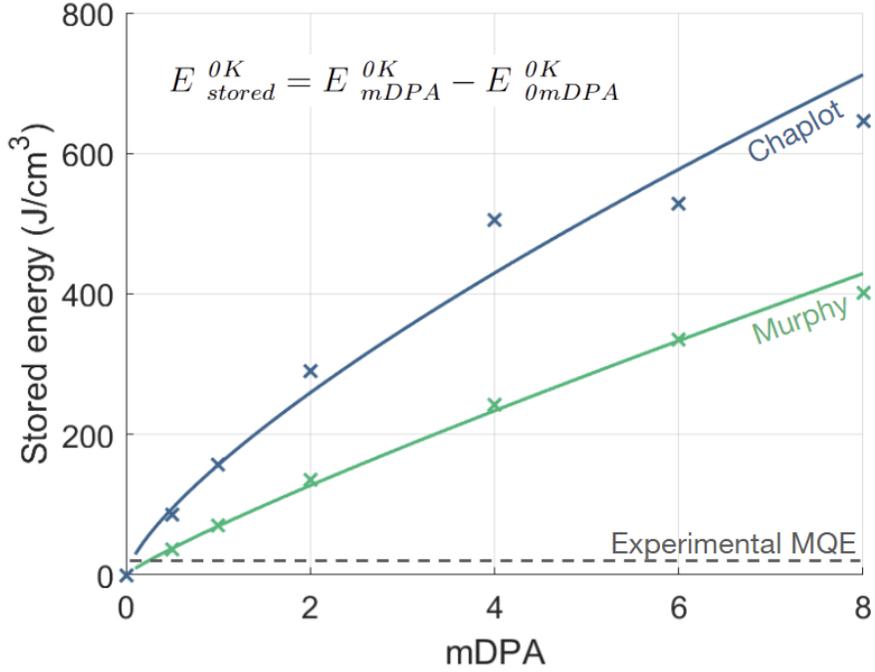

**Figure 2. The simulated stored energy for defected YBCO at 0K is greater than experimental MQE values.** Each datapoint is the mean of eight independent simulations, and the maximum standard deviations for each potential occur at 8 mDPA where the values are $646 \pm 12.8 \, \text{J/cm}^3$ (Chaplot) and $402 \pm 2.42 \, \text{J/cm}^3$ (Murphy).

As might be expected, the energy stored in defects increases with irradiation dose for both YBCO potentials. The data is fit to a power law, represented by the solid lines in the figure, resulting in the following equations:

$$Murphy: \quad E_{stored} = 69.3 \cdot mDPA^{\,0.876} \tag{3}$$

$$Chaplot: \quad E_{stored} = 156 \cdot mDPA^{\,0.729} \tag{4}$$

The Murphy potential [22] yields stored energy values that are 55% smaller than those produced by the Chaplot potential [21]. Building on the work of Chaplot (which targeted equilibrium properties), Murphy and Gray [36] allowed the exchange of Cu ions between two symmetrically distinct sites to prevent artificial charge distortion.

This provides more pathways for defect relaxation, and could contribute to the lower stored energy values. Notably, the energy stored in defects is much greater than the experimental MQE values from literature [1, 14]. However, one should exercise caution when comparing between these values as the simulations were conducted at 0 K, and all experimental measurements were made at non-zero temperatures. To correct for this artifact, the effect of temperature on the evolution of stored energy was explored through isothermal annealing simulations, where supercells were relaxed at temperatures of 1 K, 20 K, 40 K, 92 K, and 300 K for 35 ps.

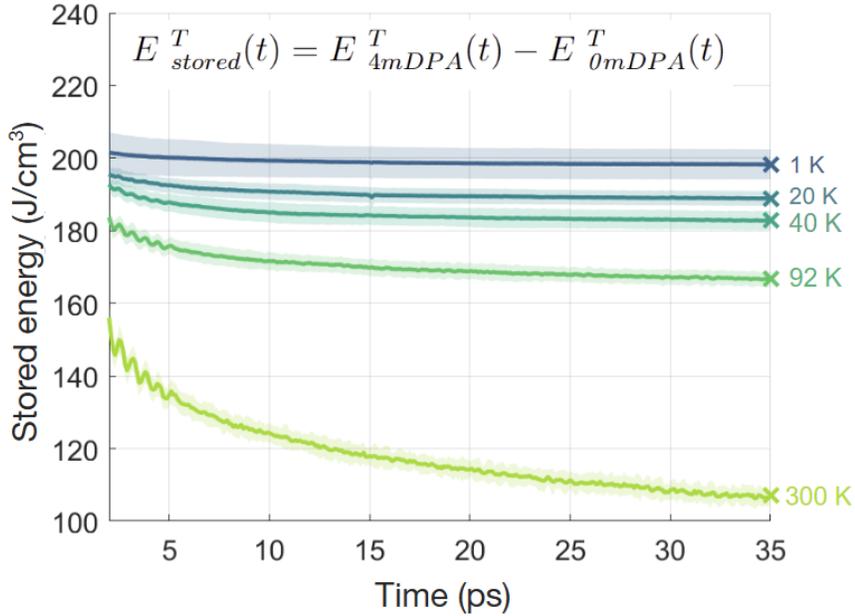

**Figure 3. Isothermal annealing at higher temperatures leads to lower values of stored energy.** This data is the result of 3 independent simulations. The shaded regions represent the standard deviation, while the lines represent the average of the simulations.

Figure 3 illustrates the results of these simulations, for the 4 mDPA case, where higher isothermal relaxation temperatures leads to a greater decrease in the stored energy over time. Higher temperatures increase the diffusivity of atoms, allowing defects to migrate further in a given time, increasing the likelihood of interaction with other defects. Recombination and clustering both lead to more favorable atomic configurations, reducing stored energy in the material. Note that the oscillations observed in the first 5 ps of the data are an artifact due to the equilibration period of the thermostat process, where particle velocities are initiated based on the designated temperature.

At 300 K, the stored energy of an 4 mDPA cell decreased by approximately 28%, whereas at 20 K (the proposed operating temperature for fusion magnets [8]) the stored energy decreased by only 1% during the simulation time. The simulation results also show that the rate of energy release decreases as a function of time. This rate may be intrinsically related to the defect density and associated defect-defect spacing. Initially, the rapid energy release may be attributed to the recombination of correlated Frenkel pairs, and as these defects recover, longer range defect migration must occur before subsequent clustering or annihilation.

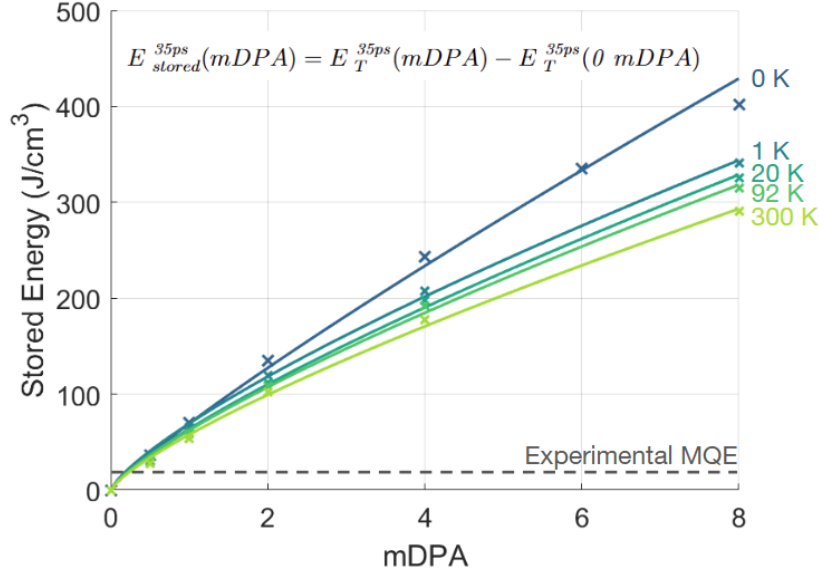

**Figure 4. The simulated stored energy for defected YBCO (at T>0 K) is greater than the experimental MQE values for doses greater than 0.1 mDPA.** Each point is the average of three independent simulations, and the standard deviation is smaller than the marker size. The highest standard deviation occurs at 300 K and 4 mDPA, 177 ± 2.46 J/cm$^3$.

| Temperature (K) | $E_{stored} = A \cdot (mDPA)^n$ | | $R^2$ |
|---|---|---|---|
| | $A$ | $n$ | |
| 1 | 69.4 | 0.770 | 0.999 |
| 20 | 63.8 | 0.788 | 0.998 |
| 40 | 62.2 | 0.785 | 0.998 |
| 92 | 57.8 | 0.781 | 0.998 |
| 300 | 37.7 | 0.814 | 0.997 |

**Table 1.** Power-law fits to the data of stored energy as a function of irradiation dose for different temperatures. Note that the 40 K data is not shown in Figure 4 for clarity.

The values of stored energy remaining at 35 ps, shown as the (×) markers in Figure 2, are plotted in Figure 5 for all irradiation doses. These values represent the potential energy difference between defected and perfect supercells annealed at the same temperature, isolating the energy contribution from surviving defects. At elevated temperatures, increased defect diffusion facilitates recombination, further reducing stored energy. The post-annealing stored energy values correct for the thermal recombination of defects, which is not captured by the FPA algorithm at 0 K. Magnet operating temperatures will ultimately dictate the maximum amount of stored energy retained as a function of dose. Even at elevated temperatures, the stored energy remains higher than experimental MQE values, which emphasizes the need to investigate this phenomenon further, including longer timescale simulations and experimental validation of our results.

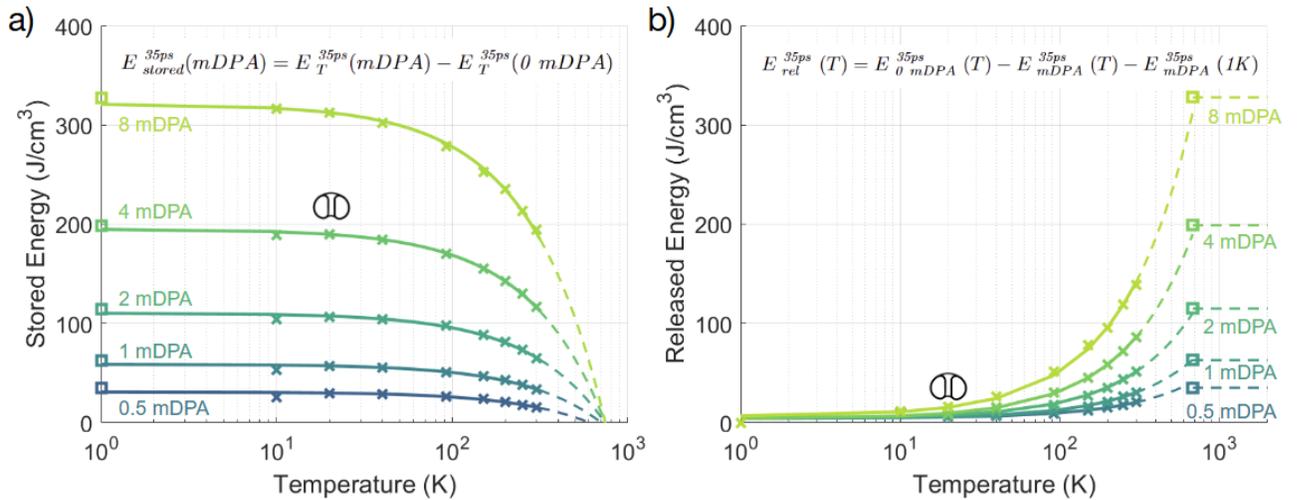

Figure 5. **Stored energy decreases and released energy increases with increasing annealing temperature.** Plots showing the stored (a) and released (b) energy after 35 ps annealing simulations at temperatures from 1–300 K with different starting defect concentrations of 0.5—8 mDPA. This data is the result of 3 independent simulations, and the standard deviation is smaller than the marker size. The highest standard deviation occurs at 300 K and 4 mDPA, with a value of $177 \pm 2.46$ J/cm$^3$.

The isothermally-relaxed stored energy values from Figure 4 are presented in Figure 5 as a function of temperature. Extrapolating a linear fit across the stored energy (a) plot yields an interesting convergence of all doses at 680 K $\pm$ 10 K, which is approximately half of the 1300 K YBCO melting point [37]. The released energy trend shown in (b) is linear (note: both graph axes are logarithmic) with temperature. However, these values cannot be extrapolated indefinitely as the maximum value of released energy cannot exceed the 0 K stored energy at a given dose, which is marked by square datapoints in the released energy plot, Figure 5(b). A tokamak icon is used to denote the intended toroidal field magnet operating temperature for YBCO in a fusion reactor and the estimated lifetime dose of 4 mDPA [12].

## 4. Discussion

Molecular dynamics simulations show that YBCO could store sufficient energy, in the form of lattice defects, to accelerate, or possibly trigger, a magnet quench. In the case of a magnet at the end of its operational life ($\sim$ 4 mDPA, 20 K), our simulations show that heating YBCO to 92 K could release up to 20 J/cm$^3$. Considering an experimental MQE in the range of 1.4–8.1 J/cm$^3$ [1, 14] (data from YBCO tape quench experiments at 77 K in self-field), the risk of energy released by defect annealing contributing to a magnet quench may be significant.

Our initial set of 0 K calculations (Figure 2) show that the energy stored in irradiation-induced defects can reach values of several hundred J/cm$^3$, and these values surpass experimentally measured MQE values above doses of $\sim$0.25 mDPA. This result holds for both interatomic potentials tested, which demonstrates that the findings are independent of the chosen potential. The calculated YBCO defect stored energies (300-400 J/cm$^3$) are comparable to those computed [38,39] for other materials such as silicon (400 J/cm$^3$), graphite ($\sim$1750 J/cm$^3$, 0.1 at.%) and diamond (4900 J/cm$^3$), noting that YBCO harbors covalent and ionic bonds. Figure 2 also shows that defect concentration does not saturate within the dose range relevant to fusion magnet operations ($< 8$ mpda), meaning that our calculated stored energy values may be lower than the theoretical maximum value. However, as stated above, these calculations were conducted at 0 K and the effect of irradiation at a finite temperature must be included.

The time evolution of stored energy (Figure 3) shows that the rate of energy release decreases significantly within 5 ps, after which the system enters a regime of slow, gradual energy release. The implications of this result are two-fold. First, significant potential energy may be retained in the superconductor after relaxation. Second, although short simulations (e.g. 35 ps, 300 CPU-hours) are sufficient to estimate the majority of energy uptake, they may not capture slower recombination and clustering processes that continue to reduce the stored energy over longer timescales. In contrast, the 300 K data exhibits a higher rate of energy release at 35 ps, indicating the importance of extended simulations at elevated temperatures. This limitation is particularly relevant given that quench events in HTS magnets evolve over timescales ranging from milliseconds to seconds [40], 9 orders of magnitude longer than the simulations presented here. As a result, defect dynamics observed at early times may represent only the initial phase of a much longer energy release process. The observed trend has a rapid initial decrease in stored energy that gradually slows over time, resembling power-law decay behavior. This decay may be attributed to the diffusion-controlled random walk of defects driving recombination and the accompanying reduction in defect density.

In YBCO, the most energetically favorable and commonly observed radiation-induced defects are oxygen vacancies and copper interstitials, which have been observed in both DFT simulations [22] and experiments [41]. Oxygen vacancies have a low formation energy (0.88-1.71 eV) and weak binding in the Cu-O chains within the lattice, leading to preferential formation during a damage cascade [36]. Copper interstitials are more likely to form than yttrium due to their smaller atomic radius and higher mobility. While the simulations in this work used randomized Frenkel pair insertion to introduce defects - an approach that provides statistical coverage across species and lattice sites - it does not fully reflect the preferential formation of specific defect types observed experimentally and in DFT studies. Still, this method offers a useful approximation for estimating the upper bound of defect-stored energy and capturing general trends in defect behavior. The dynamics of defect evolution in YBCO are highly species dependent. Even at cryogenic temperatures, oxygen interstitials and vacancies are highly mobile, leading to fast recombination and short-lived defect lifetimes [36]. In contrast, larger species such as copper and yttrium exhibit lower diffusivities and longer lifetimes. The atomic dynamics are further complicated by YBCO's anisotropic and layered structure. The diffusion of oxygen is more favored along Cu-O chains than in the $CuO_2$ planes or the Ba-O layers due to the differences in structural and bonding environments [42]. Given the picosecond timescales accessible to molecular dynamics, our simulations likely capture the fast behavior of interstitial-driven recombination, while slower processes such as defect clustering, long-range diffusion, and annihilation at grain boundaries unfold on nanosecond to millisecond timescales and are the subject of future work.

The energy stored in defects at different temperatures (Figure 4) shows that while temperature does affect the magnitude of stored energy, the dominant factor in determining stored energy on short timescales is dose. The power law fits to stored energy data have approximately the same exponents, with values ranging from 0.770 to 0.814 as temperature increases, suggesting that the fundamental scaling behavior with dose is robust. Meanwhile, the prefactor is temperature dependent, decreasing from 69.4 to 37.7 as temperature increases from 1 K to 300 K. Higher temperatures increase defect mobility, leading to faster recombination and an associated reduction in stored energy. Although our simulations show relative energy release consistent with increased diffusion-driven recombination at elevated temperatures, explicitly confirming the Arrhenius behavior would require temperature-dependent diffusivity calculations, which were beyond the scope of this study.

The energy stored in defects (Figure 5) follows a linear trend across temperatures, consistent with a diffusion-limited process where the recombination rate is controlled by the mobility of defects. As temperature increases, the diffusivity of interstitials increases, resulting in more frequent interaction with vacancies. In addition, we cannot exclude other thermally activated mechanisms such as vacancy or interstitial clustering, which may become increasingly significant at higher temperatures and longer timescales. Extrapolating to higher temperatures reveals an intriguing intersection where all lines converge to 0 J/cm$^3$ at 680 K. This convergence, independent of dose, suggests that the model reaches an equivalent energy state at 680 K, where temperatures in excess of 680 K have released all defect stored energy. This saturation is denoted as the square datapoints in subfigure b, with values equal to the stored energy at 1 K. These values also represent the maximum amount of energy that could be released via defects. It should be noted that the defects simulated using FPA are isolated interstitials and vacancies, and the energy release observed likely arises from recombination of these isolated defects. Under neutron irradiation, intracascade clustering will be more pronounced, leading to the athermal formation of defect clusters which will have lower mobilities than isolated defects. As a result, our simulations represent an upper limit on the recombination rate, and thus energy release rate.

Considering the real-world consequences of defect stored energy in YBCO magnets, it is important to note the significant difference between MD simulations of an infinite YBCO single crystal, and experiments on polycrystalline, multilayered HTS tapes and magnet cables. Experimental MQE values depend on the operating temperature, applied current, magnetic field, and structural properties. Commercially-produced YBCO tapes contain grain boundaries and $Y_2O_3$ nanoparticle artificial pinning centers [13], which alter both the thermal and electrical transport as compared to a pristine single crystal; leading to a reduced overall MQE [8]. In addition, the total energy stored and subsequently released will be affected by the presence of grain boundaries. Grain boundaries act as defect sinks, during irradiation and immediately afterwards, leading to a reduction in the saturation defect density of the material. On a macro-scale level, experiments have shown that the twisting and stacking configurations of HTS tapes into cables affect MQE [43]. These factors combine to prevent the direct comparison between our simulations with practical scenarios, without extensive modeling of the many micro-, meso-, milli-structural features that that exist between them. Our quantified energy release should not be directly interpreted as MQE, but instead as one of the many contributing factors to a quench. The timescales and irradiation conditions modeled in our simulations differ significantly from those in experiments; our simulations probe defect behavior over tens of picoseconds, but real quench events unfold over milliseconds to seconds. Similarly, our method of defect insertion does not replicate the behavior of damage accumulation from neutron fluence or ion irradiation, and the simulated dose does not directly correspond to an experimental dpa or fluence value. As a result, the most expeditious route towards understanding this phenomenon may instead be the *experimental* investigation of stored defect energy in irradiated YBCO magnet materials.

There are many avenues for further work, including the following:

(i) The defect configurations investigated in this study, isolated Frenkel pairs, are most comparable to electron irradiation at 0 K. An initial defected microstructure that more closely resembles the expected neutron radiation damage could be created through the use of consecutive displacement cascade simulations [30];

(ii) Insight into the atomistic mechanisms behind the accumulation and release of stored energy should be gained through visualization and analysis of the output atomic configurations;

(iii) Simulation durations could be extended beyond MD to capture long-timescale recovery mechanisms by using alternative techniques such as Self-Evolving Atomistic Kinetic Monte Carlo (SEAKMC) simulations [44]. While MD can capture fast processes like interstitial migration, it is limited by short simulation timescales and may miss slower mechanisms such as vacancy migration and defect clustering. These alternative simulation methods also allow for the determination of diffusion coefficients, recombination rates, and stored energy release over annealing timescales more relevant to experimental and reactor conditions;

(iv) The stored energy accumulation and release simulations shown in this study should also be conducted on the other materials present in YBCO tapes, including copper (20 $\mu$m), silver (2 $\mu$m), Hastelloy (40 $\mu$m), and oxide buffer layers (0.5 $\mu$m) [45], defect accumulation, annihilation, and heat propagation in other layers could affect evolution within the YBCO layer;

(v) Validation of the 0 K stored defect energy values could be conducted using a DFT-driven Creation-Relaxation Algorithm (CRA) [46] to obtain independent, ab-initio defect energy estimates. Unlike molecular dynamics simulations based on interatomic potentials, DFT-CRA does not rely on a fitted potential and can serve as a benchmark for assessing the accuracy of the Murphy and Chaplot potentials in predicting defect formation energies;

(vi) Using the stored energy release values from all HTS layers, a multiphysics simulation of a quench could be developed to observe dynamics on a larger scale;

(vii) MQE is an important benchmark, and it would be beneficial to experimentally measure the MQE of (irradiated) single crystal YBCO to have a closer comparison to the simulated released energy values;

(viii) *In situ* cryogenic irradiation calorimetry experiments should be conducted to verify the results of our simulations and thus quantify the magnitude and kinetics of stored defect energy release in YBCO.

## 5. Conclusion

Molecular dynamics simulations of point defect accumulation and annihilation in YBCO have been used to investigate the balance between potential energy storage, and release, with respect to the minimum quench energy of superconducting magnets. Our simulations demonstrate that the energy stored in crystalline defects is significantly greater than experimental minimum quench energy values for YBCO tapes. These findings imply that irradiated YBCO could contain sufficient energy which, if released, could locally heat the material, accelerating or even triggering a magnet quench. Our results show that the operating temperature and irradiation dose both have a significant impact on the energy release during annealing. The simulations indicate that stored energy values follow a power-law dependence on dose, and the energy release follows a diffusion-limited trend with temperature, converging toward full recombination around 680 K. This study demonstrates the critical significance of further work into recovery mechanisms within irradiated YBCO, through longer timescale simulations and their experimental validation. This work should be expanded upon to accurately quantify the magnitude and, crucially, the release kinetics of stored defect energy in irradiated YBCO. The energy stored in defects may have significant implications for the reliable operation of superconducting magnets in fusion power plants, and must be investigated further.

## 6. Author Contributions

**Lauryn Kortman:** Data Curation, Writing – Original Draft, Formal Analysis, Methodology, Software, Visualization, Investigation. **Alexis Devitre:** Conceptualization, Supervision, Writing – Review & Editing, Investigation. **Charles Hirst:** Conceptualization, Methodology, Supervision, Writing – Review & Editing, Investigation.

## 7. Data Availability

All LAMMPS code to generate data and MATLAB code to graph the data presented in this paper can be accessed here: https://github.com/hirstlab/YBCO-stored-energy

## 8. Acknowledgements

The authors would like to thank the sponsors of the inaugural Plasma Science and Fusion Center (PSFC) Fusion Undergraduate Scholars (FUSars) program for supporting undergraduate fusion energy research, and Prof. M. Short and Ms. R. Shulman for implementing the FUSars program with a successful first run! We would also like to acknowledge the MIT SuperCloud and Lincoln Laboratory Supercomputing Center for providing HPC and consultation resources that have contributed to the research results reported within this paper. Special thanks are addressed to Mr. Burns for supporting nuclear research at MIT!

## 9. References

[1] Yukikazu Iwasa. *Case Studies in Superconducting Magnets*. Springer.
[2] D. G. Whyte, J. Minervini, B. LaBombard, E. Marmar, L. Bromberg, and M. Greenwald. Smaller sooner: Exploiting high magnetic fields from new superconductors for a more attractive fusion energy development path. *Journal of Fusion Energy*, 35(1):41–53, Jan 2016.
[3] B. N. Sorbom, J. Ball, T. R. Palmer, F. J. Mangiarotti, J.M. Sierchio, P. Bonoli, C. Kasten, D. A. Sutherland, H. S. Barnard, C. B. Haakonsen, J. Goh, C. Sung, and D. G. Whyte. The engineering design of ARC: A compact, highfield, fusion nuclear science facility and demonstration power plant. In *2015 IEEE 26th Symposium on Fusion Engineering (SOFE)*, pages 1–6, Austin, TX, USA, May 2015. IEEE.
[4] G Rutherford, H S Wilson, A Saltzman, D Arnold, J L Ball, S Benjamin, R Bielajew, N de Boucaud, M Calvo-Carrera, R Chandra, and et al. Manta: A negative-triangularity nasem-compliant fusion pilot plant. *Plasma Physics and Controlled Fusion*, 66(10):105006, Aug 2024.
[5] D X Fischer, R Prokopec, J Emhofer, and M Eisterer. The effect of fast neutron irradiation on the superconducting properties of REBCO coated conductors with and without artificial pinning centers. *Superconductor Science and Technology*, 31(4):044006, April 2018.

[6] A. L. Solovjov, E. V. Petrenko, L. V. Omelchenko, R. V. Vovk, I. L. Goulatis, and A. Chroneos. Effect of annealing on a pseudogap state in untwinned YBa2Cu3O7 single crystals. *Scientific Reports*, 9(1):9274, June 2019.

[7] Raphael Unterrainer, David X Fischer, Alena Lorenz, and Michael Eisterer. Recovering the performance of irradiated high-temperature superconductors for use in fusion magnets. *Superconductor Science and Technology*, 35(4):04LT01, April 2022.

[8] Erica Salazar. *Quench dynamics and fiber optic quench detection of VIPER high temperature superconductor cable*. Thesis, Massachusetts Institute of Technology, September 2021.

[9] E. P. Wigner. Theoretical physics in the metallurgical laboratory of chicago. In Alvin M. Weinberg, editor, *Nuclear Energy*, pages 452–458. Springer Berlin Heidelberg, Berlin, Heidelberg, 1992.

[10] S Foltyn, L Civale, Judith MacManus-Driscoll, Quanxi Jia, Boris Maiorov, Haiyan Wang, and M Maley. Materials science challenges for high-temperature superconducting wire. *Nature materials*, 6:631–42, 10 2007.

[11] Lance L. Snead, Yutai Katoh, Takaaki Koyanagi, and Kurt Terrani. Stored energy release in neutron irradiated silicon carbide. *Journal of Nuclear Materials*, 514:181–188, February 2019.

[12] Zachary S. Hartwig, Rui F. Vieira, Brandon N. Sorbom, Rodney A. Badcock, Marta Bajko, William K. Beck, Bernardo Castaldo, Christopher L. Craighill, Michael Davies, Jose Estrada, Vincent Fry, Theodore Golfinopoulos, Amanda E. Hubbard, James H. Irby, Sergey Kuznetsov, Christopher J. Lammi, Philip C. Michael, Theodore Mouratidis, Richard A. Murray, Andrew T. Pfeiffer, Samuel Z. Pierson, Alexi Radovinsky, Michael D. Rowell, Erica E. Salazar, Michael Segal, Peter W. Stahle, Makoto Takayasu, Thomas L. Toland, and Lihua Zhou. VIPER: an industrially scalable high-current high-temperature superconductor cable. *Superconductor Science and Technology*, 33(11):11LT01, October 2020.

[13] A. Molodyk, S. Samoilenkov, A. Markelov, P. Degtyarenko, S. Lee, V. Petrykin, M. Gaifullin, A. Mankevich, A. Vavilov, B. Sorbom, J. Cheng, S. Garberg, L. Kesler, Z. Hartwig, S. Gavrilkin, A. Tsvetkov, T. Okada, S. Awaji, D. Abraimov, A. Francis, G. Bradford, D. Larbalestier, C. Senatore, M. Bonura, A. E. Pantoja, S. C. Wimbush, N. M. Strickland, and A. Vasiliev. Development and large volume production of extremely high current density YBa2Cu3O7 superconducting wires for fusion. *Scientific Reports*, 11(1):1–11, 2021.

[14] Jiangtao Shi, Tian Qiu, Wei Chen, Haiyang Zhang, Xinsheng Yang, and Yong Zhao. Effect of the local defects induced by bending strain on the quench properties for YBCO tapes. *Cryogenics*, 90:52–55, March 2018.

[15] W.J. Lu, J. Fang, D. Li, C.Y. Wu, and L.J. Guo. The experimental research and analysis on the quench propagation of YBCO coated conductor and coil. *Physica C: Superconductivity*, 484:153–158, January 2013.

[16] Ram Gopal Sharma. *Superconductivity: Basics and Applications to Magnets*, volume 214. Springer, 2015.

[17] G. P. Summers, E. A. Burke, D. B. Chrisey, M. Nastasi, and J. R. Tesmer. Effect of particle-induced displacements on the critical temperature of yba$_2$cu$_3$o$_{7\delta}$. *Applied Physics Letters*, 55(14):1469–1471, 1989.

[18] Luca Reali, Mark R. Gilbert, Max Boleininger, and Sergei L. Dudarev. -photons and high-energy electrons produced by neutron irradiation in nuclear materials. *Journal of Nuclear Materials*, 585:154584, 2023.

[19] M.A Kirk and Y Yan. Structure and properties of irradiation defects in yba 2 cu 3 o 7x. *Micron*, 30(5):507–526, Oct 1999.

[20] F Z Cui, J Xie, and H D Li. Preferential radiation damage of the oxygen sublattice inyba2cu3. *Physical review. B, Condensed matter*, 46(17):11182–11185, Nov 1992.

[21] S. L. Chaplot. Molecular-dynamics simulation of YBa $_2$ Cu $_3$ O $_7$ at high temperatures. *Phase Transitions*, 19(1-3):49–59, November 1989.

[22] Samuel T Murphy. A point defect model for YBa $_2$ Cu $_3$ O $_7$ from density functional theory. *Journal of Physics Communications*, 4(11):115003, November 2020.

[23] T. Trevethan and M.I. Heggie. Molecular dynamics simulations of irradiation defects in graphite: Single crystal mechanical and thermal properties. *Computational Materials Science*, 113:60–65, 2016.

[24] Xiu Fu, Zongwei Xu, Zhongdu He, Alexander Hartmaier, and Fengzhou Fang. Molecular dynamics simulation of silicon ion implantation into diamond and subsequent annealing. *Nuclear Instruments and Methods in Physics Research Section B: Beam Interactions with Materials and Atoms*, 450:51–55, 2019. The 23rd International Conference on Ion Beam Analysis.

[25] Aidan P. Thompson, H. Metin Aktulga, Richard Berger, Dan S. Bolintineanu, W. Michael Brown, Paul S. Crozier, Pieter J. In 'T Veld, Axel Kohlmeyer, Stan G. Moore, Trung Dac Nguyen, Ray Shan, Mark J. Stevens, Julien Tranchida, Christian Trott, and Steven J. Plimpton. LAMMPS - a flexible simulation tool for particle-based materials modeling at the atomic, meso, and continuum scales. *Computer Physics Communications*, 271:108171, February 2022.

[26] B. Hensel, B. Roas, S. Henke, R. Hopfengärtner, M. Lippert, J. P. Ströbel, M. Vildić, G. Saemann-Ischenko, and S. Klaumünzer. Ion irradiation of epitaxial yba$_2$cu$_3$o$_{7-\delta}$ films: Effects of electronic energy loss. *Phys. Rev. B*, 42:4135–4142, Sep 1990.

[27] P. M. Derlet and S. L. Dudarev. Microscopic structure of a heavily irradiated material. *Physical Review Materials*, 4(2):023605, February 2020.

[28] Daniele Torsello, Federico Ledda, Simone Sparacio, Niccolò Di Eugenio, Matteo Di Giacomo, Erik Gallo, Zachary Hartwig, Antonio Trotta, and Francesco Laviano. Radiation environment and damage of hts magnets in an arc-like reactor. *IEEE Transactions on Applied Superconductivity*, 35(5):1–6, 2025.

[29] K. Nordlund, M. Ghaly, R. S. Averback, M. Caturla, T. Diaz De La Rubia, and J. Tarus. Defect production in collision cascades in elemental semiconductors and fcc metals. *Physical Review B*, 57(13):7556–7570, April 1998.

[30] F. Granberg, D.R. Mason, and J. Byggmästar. Effect of simulation technique on the high-dose damage in tungsten. *Computational Materials Science*, 217:111902, January 2023.

[31] Jiting Tian, Hao Wang, Qijie Feng, Jian Zheng, Xiao Liu, and Wei Zhou. Heavy radiation damage in alpha zirconium at cryogenic temperature: A computational study. *Journal of Nuclear Materials*, 555:153159, November 2021.

[32] J.C. Stimac, C. Serrao, and J.K. Mason. Dependence of simulated radiation damage on crystal structure and atomic misfit in metals. *Journal of Nuclear Materials*, 585:154633, November 2023.

[33] R. J. Cava, B. Batlogg, R. B. van Dover, D. W. Murphy, S. Sunshine, T. Siegrist, J. P. Remeika, E. A. Rietman, S. Zahurak, and G. P. Espinosa. Bulk superconductivity at 91 k in single-phase oxygen-deficient perovskite ba$_2$ycu$_3$o$_{9-\delta}$. *Phys. Rev. Lett.*, 58:1676–1679, Apr 1987.

[34] Y. Iwasa, J. Jankowski, Seung yong Hahn, Haigun Lee, J. Bascunan, J. Reeves, A. Knoll, Yi-Yuan Xie, and V. Selvamanickam. Stability and quench protection of coated ybco "composite" tape. *IEEE Transactions on Applied Superconductivity*, 15(2):1683–1686, 2005.

[35] S. Plimpton. Fast parallel algorithms for short-range molecular dynamics. Technical Report SAND-91-1144, Sandia National Lab. (SNL-NM), Albuquerque, NM (United States), May 1993.

[36] R L Gray, M J D Rushton, and S T Murphy. Molecular dynamics simulations of radiation damage in YBa$_2$Cu$_3$O$_7$. *Superconductor Science and Technology*, 35(3):035010, March 2022.

[37] Fadila Taïr, Laura Carreras, Jaume Camps, Jordi Farjas, Pere Roura, Albert Calleja, Teresa Puig, and Xavier Obradors. Melting temperature of yba2cu3o7x and gdba2cu3o7x at subatmospheric partial pressure. *Journal of Alloys and Compounds*, 692:787–792, 2017.

[38] I-Te Lu and Marco Bernardi. Using defects to store energy in materials – a computational study. *Scientific Reports*, 7(1), Jun 2017.

[39] Zhi-Gang Mei, R. Ponciroli, and A. Petersen. Wigner energy in irradiated graphite: A first-principles study. *Journal of Nuclear Materials*, 563:153663, May 2022.

[40] Andrea Zappatore, W.H Fietz, R Heller, Laura Savoldi, Michael Wolf, and Roberto Zanino. A critical assessment of thermal–hydraulic modeling of hts twisted-stacked-tape cable conductors

for fusion applications. *Superconductor Science and Technology*, 32(8):084004–084004, Jul 2019.

[41]

[42] Chong Liu, Jun Zhang, Lianhong Wang, Yonghua Shu, and Jing Fan. Molecular dynamics analysis of lattice site dependent oxygen ion diffusion in yba2cu3o7: Exposing the origin of anisotropic oxygen diffusivity. *Solid State Ionics*, 232:123–128, Dec 2012.

[43] Michael S Wolf, Reinhard Heller, Walter H Fietz, and Klaus-Peter Weiss. Design and analysis of hts subsize-conductors for quench investigations towards future hts fusion magnets. 104:102980–102980, Dec 2019.

[44] Haixuan Xu, Yury N Osetskiy, and Roger E Stoller. Self-evolving atomistic kinetic monte carlo (seakmc): Fundamentals and applications. *Journal of Physics: Condensed Matter*, 24(37), 01 2012.

[45] C Barth, G Mondonico, and C Senatore. Electro-mechanical properties of REBCO coated conductors from various industrial manufacturers at 77 K, self-field and 4.2 K, 19 T. *Superconductor Science and Technology*, 28(4):045011, April 2015.

[46] Ebrahim Mansouri and Pär Olsson. Modeling of irradiation-induced microstructure evolution in fe: Impact of frenkel pair distribution. *Computational Materials Science*, 236:112852, 2024.